\documentclass[prl,reprint,amsmath,amssymb,twocolumn,showpacs]{revtex4}
\usepackage{ulem}
\usepackage{color}

\usepackage{verbatim}
\usepackage{graphicx}
\usepackage{bm}
\def\pmb#1{\setbox0=\hbox{#1}
\kern-.025em\copy0\kern-\wd0 \kern-.05em\copy0\kern-\wd0
\kern-.025em\raise.0433em\box0}

\newcommand{\beq}{\begin{equation}}
\newcommand{\eeq}{\end{equation}}
\newcommand{\ba}{\begin{eqnarray}}
\newcommand{\ea}{\end{eqnarray}}

\input epsf

\newif\iffigures

\figurestrue 

\begin{document}

\title[]{Unveiling Extreme Anisotropy in Elastic Structured Media}

\author{G. Lefebvre$^{1}$, T. Antonakakis$^{2}$, Y. Achaoui$^{3}$, R.~V. Craster$^{4}$, S. Guenneau$^{3}$, P. Sebbah$^{1,5}$}
\affiliation{$^1$ Institut Langevin, ESPCI ParisTech CNRS UMR7587, 1 rue Jussieu, 75238 Paris cedex 05, France}
\affiliation{$^2$ Multiwave Technologies, Switzerland} 
\affiliation{$^3$ Aix-Marseille Univ., CNRS, Centrale Marseille, Institut Fresnel, Marseille, France}
\affiliation{$^4$ Department of Mathematics, Imperial College London, London SW7 2AZ, UK}
\affiliation{$^5$ Department of Physics, The Jack and Pearl Resnick Institute for Advanced Technology,
Bar-Ilan University, Ramat-Gan 5290002, Israel}

\begin{abstract}
Periodic structures can be engineered to exhibit unique properties observed at symmetry points, such as zero group velocity, Dirac cones and saddle points; identifying these, and the nature of the associated modes, from a direct reading of the dispersion surfaces is not straightforward, especially in three-dimensions or at high frequencies when several dispersion surfaces fold back in the Brillouin zone. A recently proposed asymptotic high frequency homogenisation theory is applied to a challenging time-domain experiment with elastic waves in a pinned metallic plate.  The prediction of a narrow high-frequency spectral region where the effective medium tensor dramatically switches from positive definite to indefinite is confirmed experimentally; a small frequency shift of the pulse carrier results in
two distinct types of highly anisotropic modes. The underlying effective equation mirrors this behaviour with a change in form from elliptic to hyperbolic exemplifying the high degree of wave control available and the importance of a simple and effective predictive model.


\end{abstract}
\maketitle

The analysis of periodic media underpins advances in electronic properties, wave transport in photonics and acoustics, as well as in interference phenomena throughout many fields of physical and engineering sciences. 
High symmetry points on dispersion diagrams where the group velocity tends to zero and the density of states is high, are often associated with unusual effective properties such as slow-light and ultra-directivity \cite{sakoda01a,gralak00a} when the bands display vanishing curvature and even all-angle-negative refraction \cite{luo02a}. More recently, photonics research has focused on Dirac cones following the rise of graphene \cite{neto09a,torrent13a} as well as on saddle points in the search for hyperbolic type features \cite{belov13a,drachev13a}.

However, identifying the remarkable symmetry points and the nature of their associated mode from a direct reading of the dispersion surfaces is not straightforward. 
 An asymptotic theory which easily identifies the nature of these singularities has been developed \cite{craster10a}, that does not require extensive computation of the Bloch dispersion surfaces. Indeed, in contrast to most standard methods such as semi-analytic plane-wave  \cite{soukoulis90,johnson2001} and multipole expansions \cite{poulton2000}, numerical finite element or finite difference time domain methods \cite{zolla05a}, the so-called high frequency homogenisation (HFH) approach \cite{craster10a} is purely algebraic. Because it only focuses on the apexes of the Brillouin zone, HFH captures the essential features without intensive computation and in an intuitive manner.

The basic methodology has been validated in the microwave regime \cite{ceresoli15} but for elastic waves the tensorial equations make any interpretation and prediction extremely difficult and the situation is far more demanding. By considering thin plates, the problem becomes more tractable and several achievements have been reported recently, including broadband cloaking \cite{farhat09a,stenger12}, negative refraction \cite{bonello10} and lensing \cite{dubois13,lefebvre15a} effects which parallel effects observed earlier in electromagnetic metamaterials \cite{smith04a,ramakrishna05a}. Topical large-scale applications to seismic metamaterials \cite{brule14} and experiments using pillars \cite{pennec08a} or subwavelength resonators attached to elastic plates \cite{colombi14a} provide further motivation. The Kirchhoff-Love (KL) plate equation provides the simplest model that captures flexural waves and for structured media such as phononic crystals and metamaterials the underpinning theory is being advanced accordingly \cite{movchan07c,antonakakis12a,torrent13a,mcphedran15a}.  
It is unclear the range of validity of the KL plate theory for structured media versus using full elasticity; for homogeneous plates the KL theory is only valid for thicknesses up to around $1/20^\text{th}$ of a wavelength \cite{rose04a} and we operate outside this. Furthermore, the experimental behaviour in the time domain, and the potential for hyperbolic behaviour (which is hard to find experimentally as it is highly dependent upon frequency), all remain unexplored. The full elastic system is awkward to disentangle and, given its complexity, the simplicity of the KL theory and HFH are appealing; we carefully justify the use of the simplified model in \cite{SM}. 

In this Letter, the utility of this simplified viewpoint where the elastic wave physics is captured using the simple plate model, and the frequency-dependent anisotropy identified by the HFH asymptotic approach is 
tested in the most challenging case of vector elastic waves and guides the design of a doubly-periodic clamped elastic plate capable of exhibiting the highly directive features of both elliptic and hyperbolic points.
Such a constrained elastic plate has a zero-frequency stop band that immediately excludes any possibility of using a quasi-static homogenization method. We show that HFH theory efficiently and accurately models such a system at arbitrarily high frequencies. It predicts the conditions for the observation of a narrow high-frequency spectral region where the effective medium tensor dramatically switches from positive definite to indefinite. Using heterodyne laser interferometry, we measure the time evolution of an elastic Gaussian pulse with central frequency, $\Omega_0$, within the pinned elastic plate. At the remarkable frequencies $\Omega_0$ identified by the HFH theory, we observe highly directive emission and show how a small frequency shift of the pulse carrier results in two distinct types of highly anisotropic modes. 

\iffigures
\begin{figure}[h!]
\includegraphics[width=8.5cm]{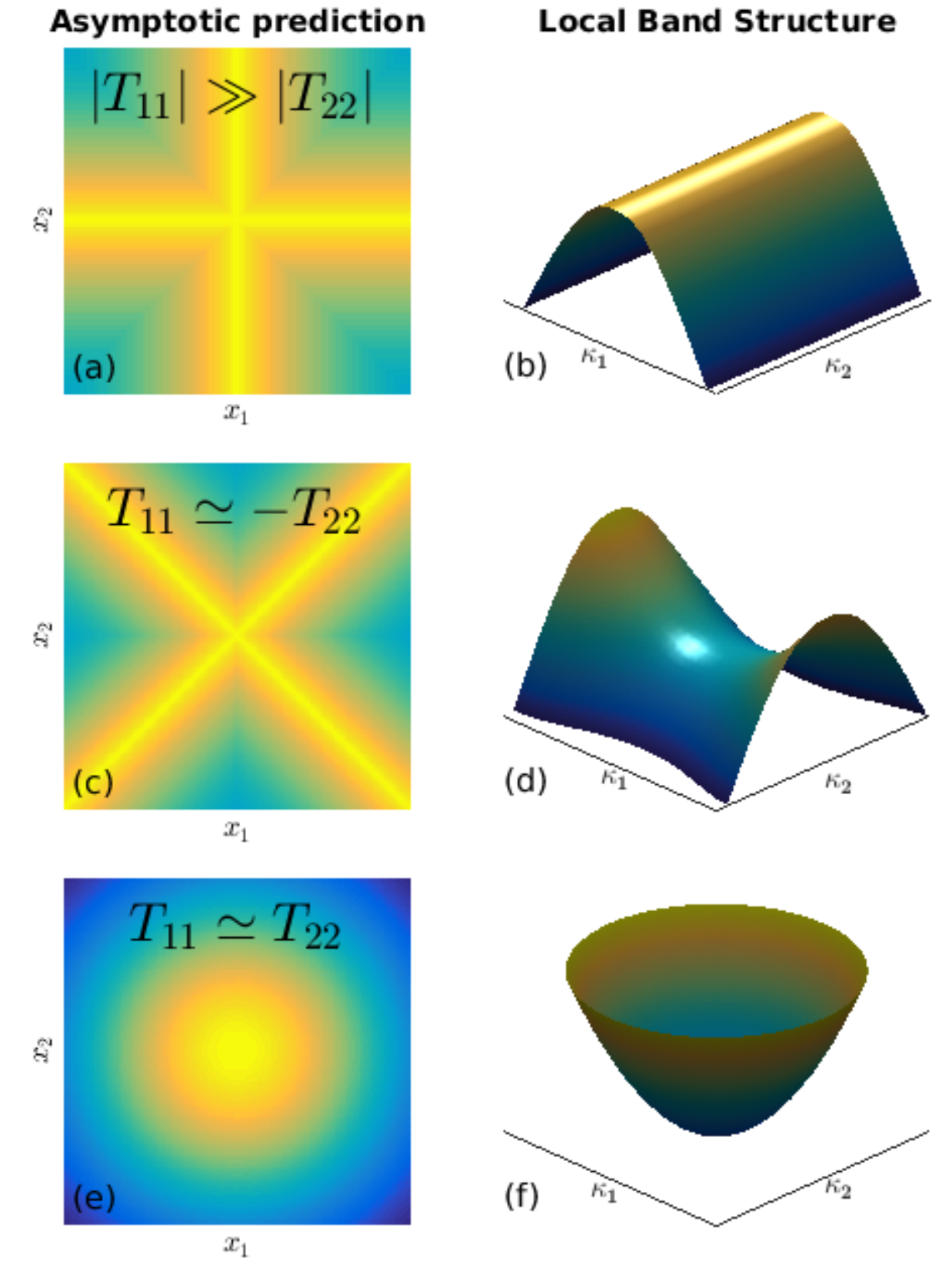} 
\caption{Asymptotic prediction of local mode structure and respective dispersion surfaces:
Effective medium tensor components govern the parabolic ($\vert T_{11}\vert\gg \vert T_{22}\vert$), hyperbolic ($T_{11}T_{22}<0$), or  elliptic ($T_{11}T_{22}>0$) character of Eq. (\ref{eq:f_0})) and capture the extreme anisotropic features of waves propagating within a doubly periodic medium shown in (a,c,e). The corresponding local dispersion surfaces 
 are shown in (b,d,f). The algebraic criteria for critical points straightforwardly extend to n-dimensional periodic systems and reveal effective elliptic ($T_{11}T_{22}...T_{nn}>0$) or hyperbolic ($T_{11}T_{22}...T_{nn}<0$) features without resorting
to visualization of hyper-surfaces.}
\label{figintro}
\end{figure}
\fi

\iffigures
\begin{figure}[h!]
\includegraphics[width=8.6cm]{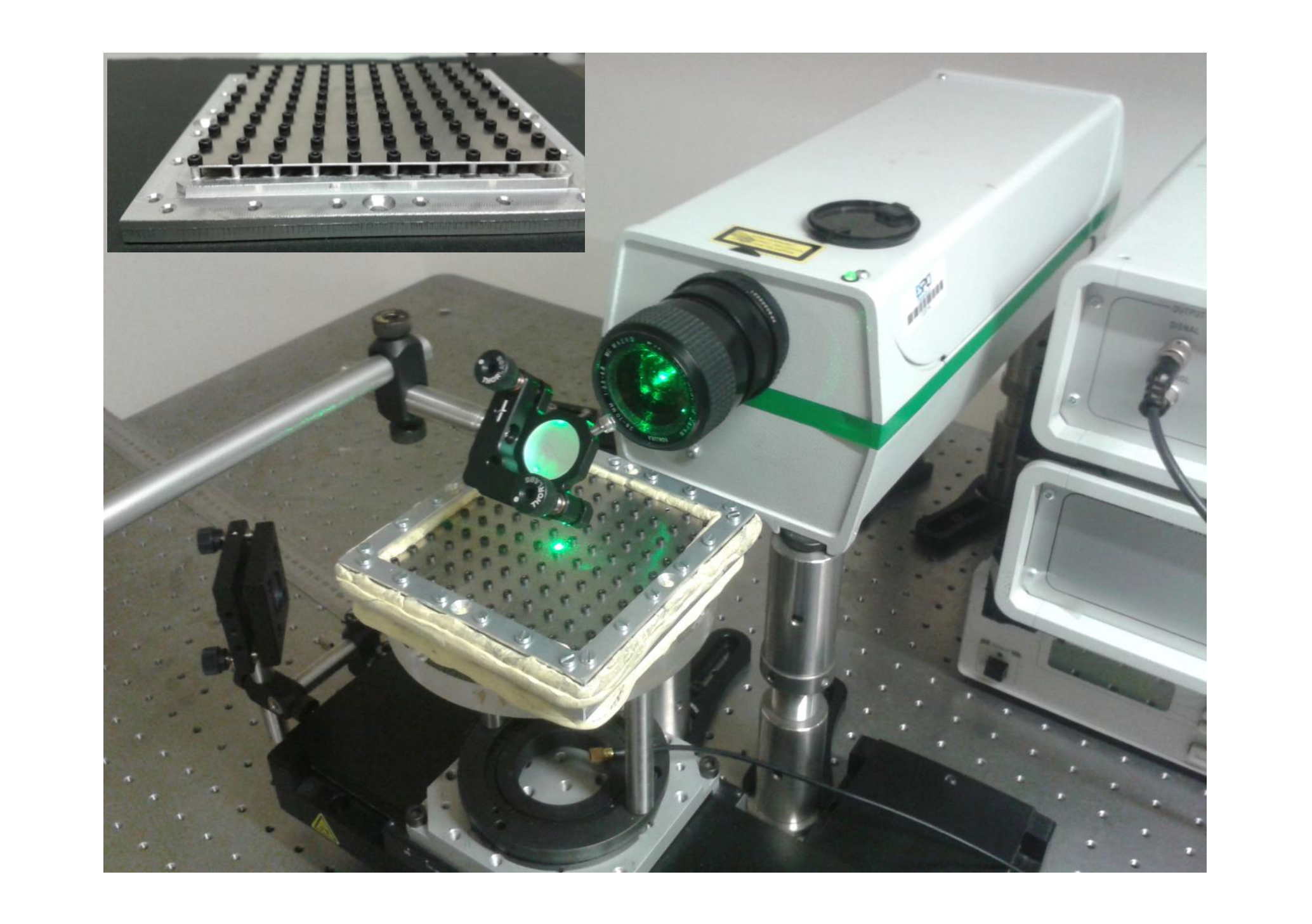}
\caption{Experimental apparatus for imaging pinned elastic plate vibrations. The heterodyne laser interferometer (Thal\`{e}s SH-140) probes the out-of-plane vibration at the surface of a 10 cm $\times$ 10 cm large, 0.5 mm-thick Duraluminium plate, clamped on a square periodic lattice by 3-mm diameter columns screwed between the thin plate and a metallic base (see inset). 3 mm-thick adhesive rubber is sandwiched between the plate's edges and a square metallic frame attached to the base, to reduce unwanted elastic edge reflections.}
\label{fig5}
\end{figure}
\fi

For waves propagating through an infinite periodic structure, one invokes Bloch's theorem \cite{kittel96a,brillouin53a} and consider an elementary cell; the Bloch wave-vector $\bm\kappa=(\kappa_1,\kappa_2)$ characterizes the phase-shift going from one cell to the next and the dispersion relation connects it with frequency. As is well-known in solid state physics \cite{brillouin53a} only a limited range of wavenumbers need normally be considered, namely the wavenumbers along the edges of the irreducible Brillouin zone. 
The HFH theory focuses on the vicinity of the vertices of the Brillouin zone, which in a square lattice reduce to 3 high symmetry points $\Gamma$, $M$ and $X$ located at the apex of a right-angled triangle. Within this framework, the wavefield, $u$, can be factorized as the product $f({\bf x})U({\bf x}; \Omega_0)$, \cite{antonakakis12a} of the Bloch solution $U({\bf x}; \Omega_0)$ at the vertex at frequency $\Omega_0$ and an envelope function $f({\bf x})$ given by the effective equation 
\beq
T_{ij}(\Omega_0)\frac{\partial^2 f}{\partial x_i\partial
  x_j}+{(\Omega_0^2-\Omega^2)}f
=0.
\label{eq:f_0}
\eeq
Eq.  (\ref{eq:f_0}) is a partial differential equation (PDE) of second order which is obtained from an asymptotic analysis of the KL equation which is fourth-order, see \cite{SM}  and \cite{antonakakis12a} for the full derivation; $\Omega$ is the wave frequency and $\Omega_0$ is the standing wave frequency at the vertex of the Brillouin zone. In essence, the factorization $f({\bf x})U({\bf x}; \Omega_0)$ captures the small scale variation of the wavefield and overcomes the limitation of standard homogenization theories, which consider the wavefield uniform at the lattice scale. At the same time, the HFH theory does not require infinite periodic structure and can describe finite-size systems.
The nature of the solutions is critically determined by the dynamic effective medium tensor $T_{ij}$. For a lattice cell with reflectional and rotational symmetry (here, a square cell with a circular clamped defect at the center), the off-diagonal terms are zero and this is just
\beq
T_{11}(\Omega_0) \frac{\partial^2 f}{\partial x_1^2}+ T_{22}(\Omega_0)\frac{\partial^2 f}{\partial x_2^2}+(\Omega_0^2-\Omega^2)f=0.
\label{eq:pde}
\eeq
We can immediately see that having both $T_{11},T_{22}$ of the same sign leads to the PDE in (\ref{eq:pde}) being of elliptic character; conversely opposite signs lead to hyperbolic behaviour. This change of character is transparently captured by the model and rapidly detected using the HFH theory. 
Similarly one can have $T_{11}\gg T_{22}$ or vice-versa leading to strong directional behaviour along one axis or the other and this behaviour too is predicted.
These three types of PDE behaviour, parabolic ($\vert T_{11}\vert \gg \vert T_{22}\vert $), hyperbolic ($T_{11}=-T_{22}$) and elliptic ($T_{11}\simeq T_{22}$), are illustrated in Fig.~\ref{figintro} where the associated modes are shown side-by-side with the corresponding local dispersion surfaces. The HFH theory therefore reduces the difficult problem of interpreting often intricate isofrequency surfaces to the straightforward comparison of the diagonal terms of the tensor, here just two numbers, $T_{ii}$.

Here we want to investigate the predictive capability of the HFH theory in the framework of elastic waves. Therefore, we consider pinned (so-called) platonic crystals (PPCs) created by an array of clamped points, with the objective of achieving highly directive emission of a periodically oscillating source located therein. PPCs are the elastic plate analogue of photonic crystals with infinite conducting rods \cite{nicorovici95a}. 
 Indeed, PPCs display a zero frequency stop band so that flexural waves with wavelengths arbitrarily long compared to the array pitch cannot propagate.

The system we want to study is shown in Fig.~\ref{fig5}. Three mm-diameter and 5 mm-high cylinders have been machined in a Duraluminium block to form a 10 by 10 square array of supports with 1 cm lattice spacing. Each support has been threaded to enable a 10 cm $\times$ 10 cm 0.5 mm-thick vibrating plate ($\rho=2789$ kg/m$^3$, E=74 GPa, $\nu$=0.33) to be attached securely to it; in the language of elastic plate theory these are clamped conditions. Although the clamped area is localised to the pin positions, the pins are not precisely point-like and so the clamping area is accounted for in the theory.

The  
KL thin plate theory is often quoted, as in say \cite{rose04a}, as only being accurate for wavelengths greater than $20$ times the plate thickness and this condition is not met at the frequencies we consider. 
However, it is the wavelength in the periodic system that actually matters, and that can be large compared to the plate thickness, even at high frequencies; we justify this technical point further by giving analogous computations for the full vector elastic system in  \cite{SM} . 


Within this thin plate framework, the explicit form of the effective medium tensor $T_{ij}$ for the PPC geometry is given in \cite{antonakakis12a}. The HFH theory is able to discriminate between interesting and uninteresting symmetry points and capture a rich array of behaviours for the given experimental parameters. In our case, we select three frequencies of interest at which modes with remarkable properties are predicted.
At 122 kHz near symmetry point X, the coefficients $T_{ii}$ in the effective tensor of the HFH theory have opposite sign, $T_{11}=-180.172$ and $T_{22}=311.7432$, and are close enough to reflect hyperbolic behaviour analog to Fig.~\ref{figintro}(b). For excitation at the center of the system, the predicted theoretical wavefield distribution at this particular frequency is shown in Fig. ~\ref{fig1}(b)): The mode is highly directive and is aligned along the diagonals of the system.
At 138 kHz, strongly differing $T_{ii}$ coefficients predict another very distinct anisotropy within the effective material (Fig.~\ref{fig1}(c)), $\vert T_{22}/T_{11}\vert \sim 0.005$ leading to the effective equation being parabolic. The mode is expected now to radiate along the lattice axis as shown in Fig.~\ref{figintro}(a). 
For completeness we discuss the situation of equal $T_{11},T_{22}$ as shown in Fig.~\ref{figintro}(c), here we expect isotropy and elliptic effective equations.
 The HFH prediction extends beyond the symmetry points with a dispersion relation local to the band edge given in  terms of the effective tensor
\beq
\Omega\sim\Omega_0-\frac{T_{ij}(\Omega_0)}{2\Omega_0}\kappa_i\kappa_j,
\label{eq:Omega}
\eeq
where $\kappa_i=K_i-d_i$ and $d_i=0,-\pi,\pi$ depending on the band edge chosen.



\iffigures
\begin{figure}[t!]
\includegraphics[width=8.0cm]{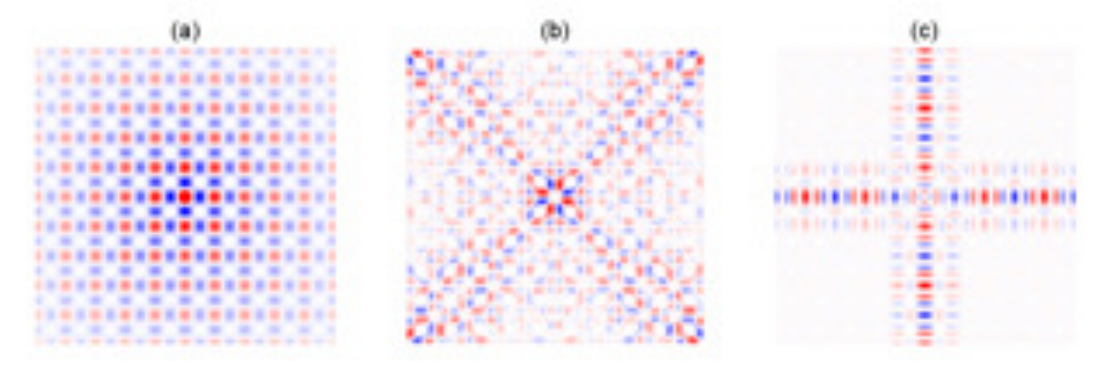}
\vspace{-2.5cm}

\includegraphics[width=8.0cm]{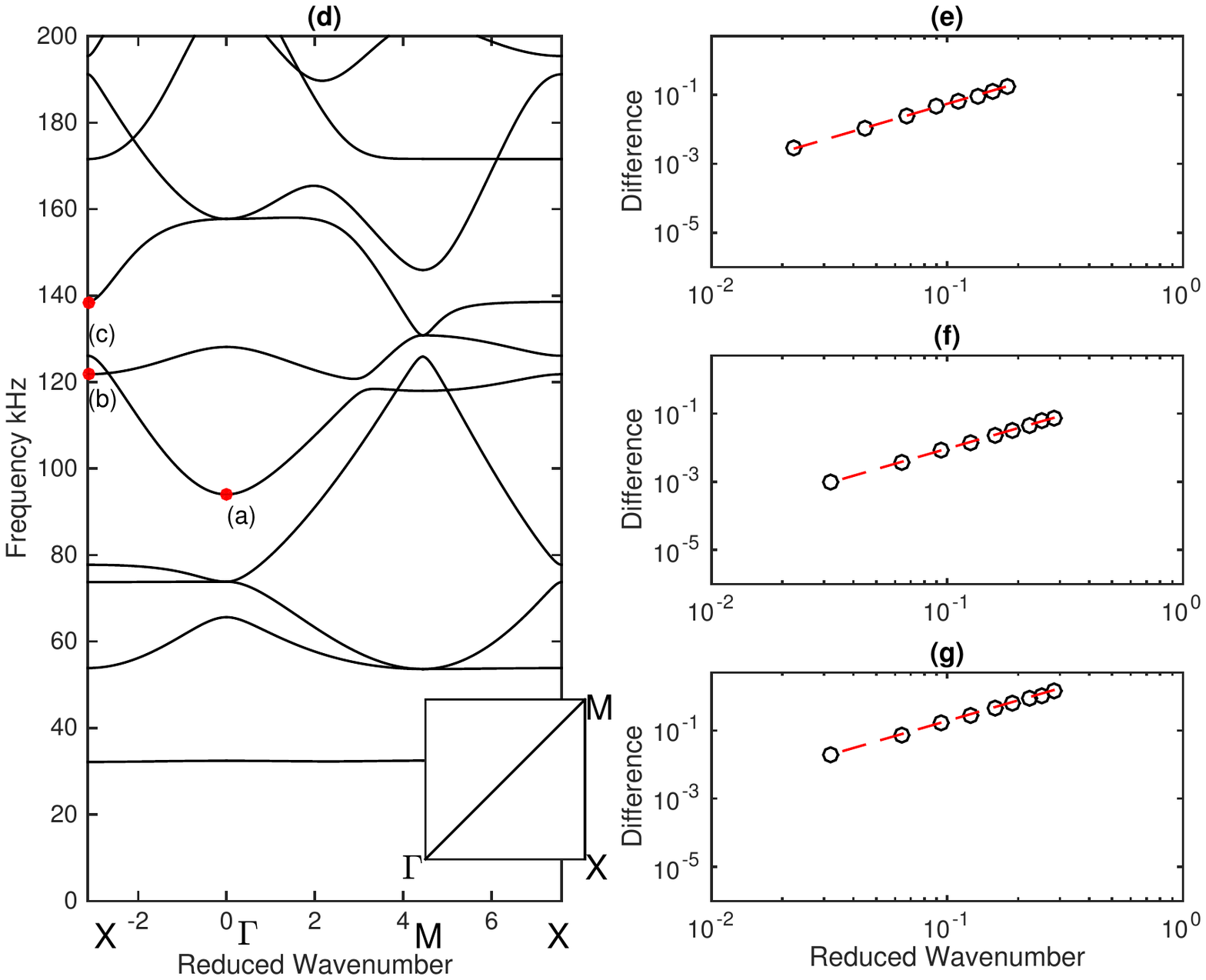}
\vspace{-2.5cm}

\caption{HFH asymptotics: (a-c) the flexural wave field predicted for a point frequency source placed centrally in the array. (a) 94 kHz, isotropic response as predicted by  effective medium tensor components $T_{11} =T_{22} = 1025$; (b)  122 kHz hyperbolic behaviour following from $T_{11} = -180.17$ and $T_{22} = 311.74$; (c) 138 kHz creating a + shape with $T_{11} = -4189.10$ and $T_{22} =  19.51$.  (d)  Bloch dispersion diagram using the irreducible Brillouin zone $\Gamma XM$ (shown inset)  with frequencies used in (a-c) as red dots. (e-g) log-log validate the HFH asymptotics in the neighbourhood of the frequencies used in (a-c). Circles are HFH asymptotics of the dispersion curves for the $X\Gamma$ path; dashed curves are from finite element numerics for the KL model and the vertical scale is the difference of frequency from that at the band edge.}
\label{fig1}
\end{figure}
\fi

In the experiment, the ultrasound emission is produced by a broadband piezoelectric transducer (Panametrics M109). A silica needle is stuck on the transducer and put in contact with the bottom surface of the plate. We generate a single-oscillation short pulse excitation synthesized by an arbitrary waveform generator (Agilent 33120A). 
The out-of-plane displacement is measured point by point with a broadband heterodyne interferometric laser probe (Thal{\`e}s SH140). The probe is scanned on a square grid with 0.68 mm step resolution and the spatio-temporal distribution of the vibration field at the surface of the plate is reconstructed. The data are subsequently filtered to obtain the elastic response to a narrow bandwidth ($\Delta\Omega_0$) Gaussian pulse with central frequency, $\Omega_0$, easily adjustable around  theoretical prediction.
To reduce elastic reflections at the edges of the plate and mimic a finite array, surrounded by perfectly matched layers, a 3 mm-thick layer of adhesive rubber is sandwiched between the edges of the vibrating plate and a rigid metallic frame (see Fig.~\ref{fig5}). This method efficiently absorbs elastic waves in the kHz range and reduces unwanted reflections \cite{lefebvre15a}.
The experimental setup is shown in Fig.~\ref{fig5}.

\iffigures

\begin{figure}[h!]
\includegraphics[width=6cm]{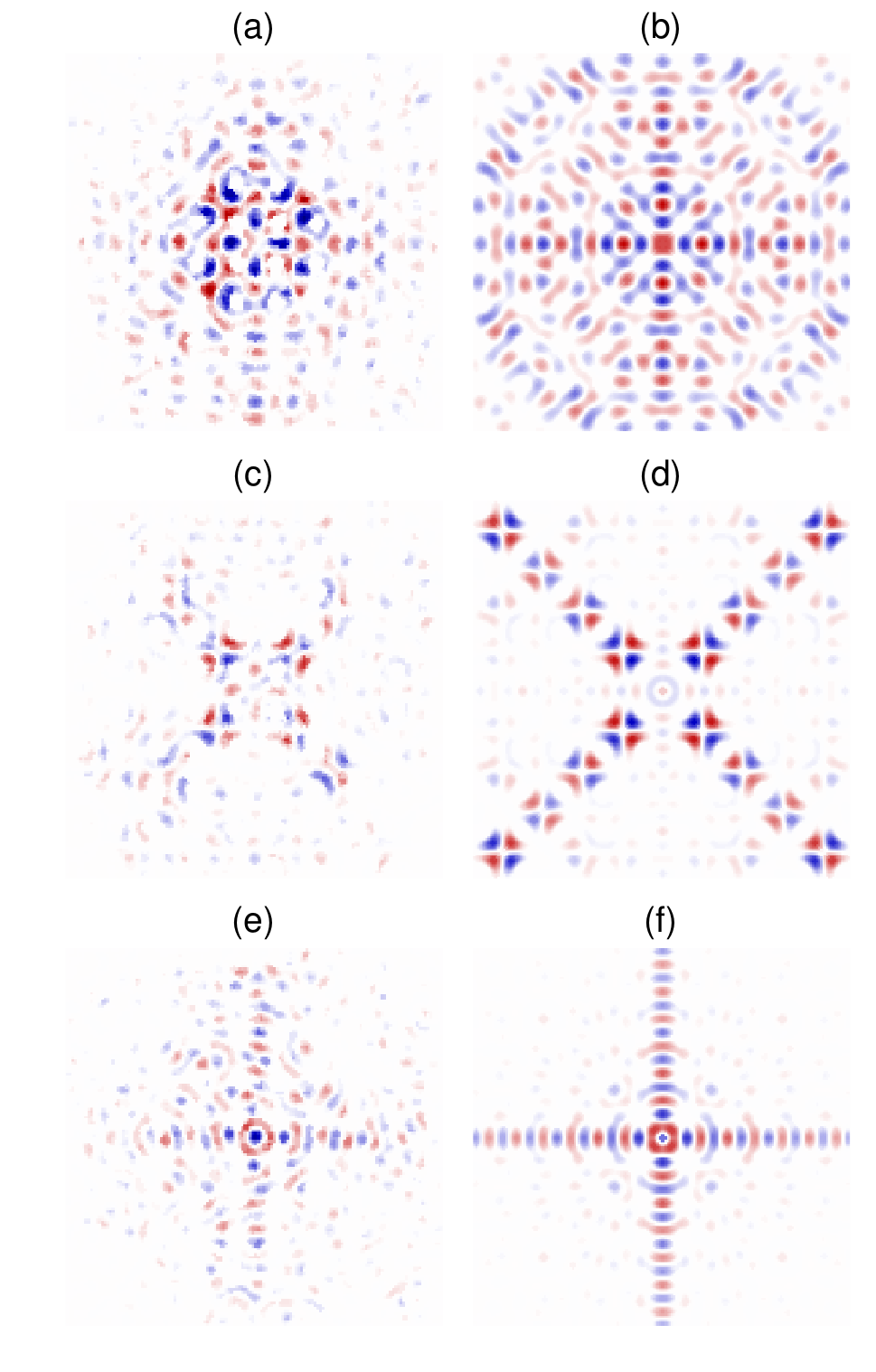}
\caption{
Snapshots of flexural wavefield evolution from a pulse excitation at the center of the system with central frequency $\Omega$ and bandwidth $\Delta\Omega$. 
 (a,c,e) experiment at $\Omega=90,105,128$ kHz and $\Delta\Omega=10$ kHz, see also the videos in
\cite{SM}. 
(b,d,f) numerical simulations, FDTD solving the full 3D Navier system, at $\Omega=95, 110, 119$ kHz with $\Delta\Omega=1$ kHz.
 The minor difference in frequency is attributed to finite 
 boundary effects in the experiments.
}
\label{fig4}
\end{figure}

\fi

The experimental results are presented in Fig.~\ref{fig4}, together with time domain simulations of the pinned-plate using the SIMSONIC package \cite{Simsonic} which solves the full 3D Navier system with the Finite Difference Time Domain (FDTD) method with absorbing boundary conditions. In contrast to the modal representation shown in Fig.~\ref{figintro}, here we  show snapshots of the time evolution of the elastic field. The complete temporal build-up of this mode is shown as a movie in the \cite{SM}. Nevertheless, the hyperbolic and elliptic behaviour are well captured with highly directional radiation at frequencies close to the prediction. Figures~\ref{fig4}c \& d show the X-shape obtained at $\Omega_0$=105 kHz ($\Delta\Omega_0$ = 10 kHz), which is characteristic of hyperbolic dispersion. This contrasts with + shape emission obtained at $\Omega_0$ = 128 kHz ($\Delta\Omega_0$ = 10 kHz) and shown in  Figs.~\ref{fig4}e \& f. The elliptic isotropic behaviour is at $\Omega_0$ = 90 kHz ($\Delta\Omega_0$ = 10 kHz) as in Figs.~\ref{fig4}a \& b. 
Resonant frequencies found numerically (95, 110 and 119 kHz) differ from the experimental frequencies by less than 10 \%. The small difference can be attributed to the  absence of residual reflections in the simulations where perfectly absorbing boundaries allow full establishment of the modes. Note also that the numerical distributions are sharper as the bandwidth used in the simulations is only $\Delta\Omega_0$ = 1kHz, leading to better spectral selectivity. Actually, the choice of the pulse bandwidth in the experiment was crucial and delicate as a trade-off was to be found between sharp modal selectivity (narrow bandwidth) and early establishment of the mode (short pulse) before edges are reached and reflections perturb the mode distribution. This problem is less of an issue in the numerics although $\Delta\Omega_0$ is kept reasonably large to limit mode establishment time and computational time.
As a result, a residual trace of the + shape mode is seen on top of the X shape mode e.g. in Fig.~\ref{fig4}d. The minor frequency offset between experiment and theory is an expected consequence of modelling the plate vibration with KL plate theory; this offset being also seen in the detailed comparisons with the full elastic dispersion diagram of \cite{SM}.

In this Letter, the recently introduced HFH was applied to the KL thin plate model, in a counter-intuitive high frequency limit where it might conventionally not be expected to hold,  to predict critical frequencies at which striking, strong anisotropic, wave patterns are generated within a periodically pinned plate. Experimental observations confirm these asymptotic predictions, as do time domain computations. 

We were drawn to this more intuitive, asymptotic approach as the available alternatives require heavy computational resources and, at high frequencies, produce results hard to interpret as illustrated in \cite{SM}. These issues are shared with many other complex doubly and triply periodic media, such as photonic and phononic crystals. The techniques developed, and illustrated herein, are valuable in these contexts.

R.V.C. thanks the EPSRC (UK) for support.  
S.G. is thankful for an ERC starting grant (ANAMORPHISM). P.S. is thankful to the Agence Nationale de la Recherche support under grant ANR PLATON (No. 12-BS09-003-01), the
LABEX WIFI (Laboratory of Excellence within the French
Program Investments for the Future) under reference
ANR-10-IDEX-0001-02 PSL* and the PICS-ALAMO. This research was supported in part by The Israel Science Foundation (Grant No. 1781/15 and 2074/15).

\bibliographystyle{unsrt}

\end{document}